\documentclass[aps,prb,twocolumn,showpacs,eqsecnum]{revtex4-1}
\usepackage{euscript}
\usepackage{amssymb}
\usepackage{amsmath}
\usepackage{graphicx}

\begin{document}

\title{Generation of electron spin polarization in disordered organic
semiconductors}

\author{A. I. Shushin}
\affiliation{Institute of Chemical Physics, Russian Academy of
Sciences, 117977, GSP-1, Kosygin str. 4, Moscow, Russia}


\begin{abstract}
The generation mechanisms of electron spin polarization (ESP) of
charge carriers (electrons and holes, called "doublets") in
doublet-doublet recombination and triplet-doublet quenching in
disordered organic semiconductors are analyzed in detail. The ESP is
assumed to result from quantum transitions between the states of the
spin Hamiltonian of the pair of interacting particles. The value of
the ESP is essentially determined by the mechanism of relative
motion of particles. In our work we have considered the cage and
free diffusion models. The effect of possible attractive
spin-independent interactions between particles is also analyzed.
Estimation with obtained formulas shows that the proposed mechanisms
can lead to a fairly strong ESP much larger than the thermal one (at
room temperatures).
\end{abstract}


\maketitle

\bigskip

\section{Introduction}

Electron spin polarization (ESP) of charge carriers (electrons and
holes) in solid-state systems is the important phenomenon, which is
observed in ferromagnetic metals and inorganic semiconductors and
result from strong magnetic interactions in these
solids.\cite{Ras,Zut} The ESP effect is very interesting from
general physical point of view and important for
applications.\cite{Ras,Zut,Tal}

As for organic semiconductors (OSCs), magnetic interactions in them
are typically fairly weak and no strong ESP is expected. The ESP
can, nevertheless, be injected from ferromagnetic solids.\cite{Tal}
In many experiments the OSCs are applied as spacers (between
ferromagnetic leads), in which one can manipulate the ESP by
applying electric and magnetic fields.\cite{Zut,Tal} The dynamics of
electron spins in OSCs is studied quite thoroughly,\cite{Zut,Tal}
but the majority of works concern the analysis of spin evolution of
polarized charge carriers injected into OSCs from polarizing leads
(see, for example, refs. [4-10], as
well as [2,3] and references therein).

This work concerns the discussion of the mechanisms of ESP
generation directly in OSCs (in the absence of spin injection). The
possibility of this effect is usually ignored because of
above-mentioned weakness of magnetic interactions in these
materials. The analysis shows, however, that some spin selective
processes in OSCs can result in fairly strong ESP (much larger than
the thermal one). Such processes are similar to those, which
determine the magnetic field and spin effects (MFEs) in spin
selective reactions of paramagnetic particles [usually doublets
($D$) and triplets ($T$)].\cite{St,Sw,Hu}

In our investigation we will consider two types of processes,
doublet-doublet ($D\!-\!D$) recombination\cite{Clos,Kap} and
triplet-doublet ($T\!-\!D$) quenching,\cite{Paul,Goud,Shu1} in which
the ESP generation is expected to be very efficient. The ESP results
from quantum transitions between the terms of the spin Hamiltonian
of the pair of particles ($D\!-\!D$ or $T\!-\!D$), assisted by their
stochastic relative motion. The terms are determined by the Zeeman
and spin exchange interactions and transitions are induced by the
hyperfine (HFI), anisotropic part of the Zeeman (AZI), or zero field
splitting (ZFS) interactions, depending on the process considered.
The strong dependence of the exchange interaction on the
interparticle distance $r$ leads to the significant localization of
transitions in space and essential dependence of the ESP generation
efficiency on the characteristic properties of relative motion.

Noteworthy is that the theory of diffusion assisted ESP generation
in liquid phase reactions has already been
developed.\cite{Goud,Shu1,Adr,Shu2,Shu3} It is based on the
stochastic Liouville equation (SLE) approach,\cite{Fre,Shu4} which
allows one to rigorously describe the generation of MFEs, (including
the ESP). As applied to similar processes in OSCs, however, this
SLE-based theory needs certain modifications, due to some specific
features of interaction and relative motion of paramagnetic
particles in these materials.

The effect of these specific features on the ESP in OSCs is
considered within two models of relative motion of particles: cage
and free diffusion. Both models are shown to predict similar
expressions for the ESP both in geminate and bulk processes. The
proposed theory enables one to take into account possible
spin-independent (i.e. Coulomb) interaction $U(r)$ between reacting
particles, which can strongly affect the MFEs.\cite{Shu5,Shu6}
Simple estimations demonstrate that the attractive interaction
$U(r)$ results in significant increase of the ESP. The evaluation of
the ESP for realistic values of parameters of the theory shows that
the considered mechanisms can be very efficient leading to a quite
large ESP in OSCs.


\section{General formulation}

The mechanisms of the ESP generation in diffusion assisted liquid
pase spin-selective reactions are thoroughly investigated in the
theory of MFEs.\cite{Goud,Shu1,Adr,Shu2,Shu3} In our work we will
briefly analyze most important results, but will be mainly concerned
with extending and adapting them to the processes under study.

We consider two types of spin-selective processes: the
doublet-doublet ($D\!-\!D$) recombination and triplet-doublet
($T\!-\!D$) quenching, in which doublet particles are associated
with electrons or holes ($D = e, h$). The most part of the
consideration concerns geminate reactions, though bulk processes are
also discussed.

The kinetic scheme the geminate variant of both processes can be
represented as
\begin{equation} \label{gen0}
X_a + X_b \stackrel{w_d^{}}\longleftarrow [X_a\dots X_b]
\stackrel{\hat K_{_R}}{\longrightarrow} X_{_R}, \;\;(X = D,T).
\end{equation}
Here $X_a$ and $X_b$ denote $D$- or $T$-particles involved in the
reaction. The reaction kinetics is essentially controlled by the
rate $w_d^{}$ of dissociation of the coupled state (cage) $[X_a\dots
X_b]$ and spin selective reaction rate $\hat K_{_R}$.

The kinetics of ESP generation in these processes is described with
the spin density matrix $\rho (t)$ of pairs of paramagnetic
particles $X_a$ and $X_b$. Naturally, $\rho (t)$ depends on the
interparticle coordinate ${\bf r}$. For simplicity, we assume that
the precess (\ref{gen0}) is isotropic, i.e. the interparticle
interactions and initial condition depend on the distance $r = |{\bf
r}|$ between particles. In this case $\rho$ depends only on $r$:
$\rho ({\bf r},t) \equiv \rho (r,t)$

The spin evolution of the pairs is governed by the spin Hamiltonian
$H (r,t)$, which can, in general, be time dependent. The explicit
form of $H (r,t)$ will be specified below for both considered
processes.

In the above-formulated assumptions the space/time evolution of the
pair density matrix $\rho (r,t)$ is determined by the SLE ($\hbar =
1$)
\begin{equation}\label{gen1}
\dot \rho = -i[H,\rho] - (\hat K_{\!_R}  + \hat {L})\rho,
\end{equation}
with $[H, \rho] = H\rho - \rho H$. In this SLE the operator $\hat L$
describes the relative motion of particles, which is treated as
hopping migration and is characterized by the average jump length
$\lambda$ and the rate of jumps $w_h$. The term $\hat K_{\!_R}\rho $
represents the effect spin selective reaction:
\begin{equation}\label{gen3}
\hat K_{\!_R}\rho =
\mbox{$\frac{1}{2}$}\kappa_{\!_R}({P}_{\!_R}^{}\rho + \rho{P}_{\!_R}^{})
\end{equation}
where ${P}_{\!_R}^{}$ is the operator of projection onto the
reactive states of the pair and $\kappa_{\!_R}$ is the reaction
rate. In general, $\kappa_{\!_R} \equiv \kappa_{\!_R} (r)$ depends
on the distance $r$ and this dependence $\kappa_{\!_R} (r)$ is
strongly localized at $r$ small distances.

The ESP essentially depends on the initial spin and spatial state.
In general, the initial density matrix can be written as
\begin{equation}\label{gen4a}
\rho (t=0)= p_0^{}(r) \rho_0^{} ,
\end{equation}
where $p_0^{}(r)$ is the initial spatial distribution (see below)
and $\rho_0^{}$ is the initial spin density matrix.

The observable under study is the ESP ${\cal P}(t \to \infty)$ of
$D$-particles. For brevity of notations, in what follow we will
mainly analyzed the absolute value of the ESP ${\cal P}$. As for the
ESP sign, it can easily be obtained, as it is demonstrated in Sec.
V.

In our work we consider the most important mechanisms of the ESP
generation in two above-mentioned processes and within cage and free
diffusion models of relative motion of particles:

1) {\it  Cage model.}\, In the cage model the considered process is
represented by the kinetic scheme (\ref{gen0}), in which $w_d^{}$ is
the rate of monomolecular dissociation of the cage and $\hat K_{_R}$
is the rate operator of spin-selective reaction, independent of the
distance $r$ ($ \kappa_{_R}$ is independent of $r$).

The operator $\hat L$, essentially controlling the kinetics of the
process [see eq. (\ref{gen3})], is defined by
\begin{equation} \label{gen7}
\hat L \rho = w_d^{} \rho.
\end{equation}

Detailed analysis shows\cite{Shu1} that non-adiabatic transitions
between terms of the Hamiltonian $H$, which determine the ESP
generation, are localized near the distance $d$ of closest approach
(of reacting particles) in the narrow region of width $\delta r \sim
\alpha^{-1} \ll d$, where $\alpha^{-1}$ is the size of the exchange
interaction (see below). The distance $d$ is, in turn, expected to
be of order of the hopping length $\lambda$. In such a case, it is
quite reasonable to assume, that the cage size $r_c \sim \lambda\;
(\sim d)$, and associate the cage dissociation rate $w_d^{}$ with
the rate $w_h$ of jumps: $w_d^{} \sim w_h^{}$.

For pairs in these cages (of volume ${\cal V}_c \sim \lambda^3$) the
relative $X_a\!-\!X_b$ coordinate ${\bf r}$ is suggested to be
randomly (homogeneously) distributed. Note that for the homogeneous
distribution over ${\bf r}$ the initial condition for the SLE can be
written in the form (\ref{gen4a}): $\rho (t = 0) = p_0^{}\rho_0$
with $p_0^{} = {\cal V}_c^{-1}$.

In the cage model, for example, the ESP of $D$-particles, escaped
from the cage, can be evaluated by formula
\begin{eqnarray} \label{gen7b}
{\cal P} &=&  2 w_d^{} \Big|\int_0^{\infty}\!\!dt\,
\big\langle {\rm Tr}[S_{_{\!D_{\!z}}}\rho(r, t)]
\big\rangle_{{\bf r} \in \,{\cal V}_c}\Big|\nonumber\\
&=& 2 w_d^{} \big|\big\langle {\rm Tr}[S_{_{\!D_{\!z}}}
\widetilde{\rho}(r,\epsilon = 0 )]
\big\rangle_{{\bf r} \in \,{\cal V}_c}\big|,
\end{eqnarray}
where $S_{_{\!D_{\!z}}}$ is the $z$-projection (along the magnetic
field ${\bf B}$) of the $D$-particle spin ${\bf S}_{_{\!D}}$,
$\tilde{\rho}(\epsilon) = \int_0^{\infty}\!dt\, e^{-\epsilon t}
\rho(t)$ is the Laplace transform of $\rho(t)$ and
\begin{equation} \label{gen7c}
\langle \rho \rangle_{{\bf r} \in \,{\cal V}_c}^{} =
{\cal V}_c^{-1}\int_{{\bf r} \in \,{\cal V}_c} \!\! d^{3}r\, \rho (r)
\end{equation}
is the average over the homogeneous distribution in the cage (of
volume ${\cal V}_c^{}$).

2) {\it Free diffusion model.}\, In the free diffusion model the
considered processes can, formally, be expressed by the same scheme
(\ref{gen0}), in which, however, the reaction and dissociation
kinetics is non-exponential and controlled by relative diffusion of
reacting particles. In this model
\begin{equation}\label{gen8}
\hat {L}\rho = -D_{\!p\,}^{} r^{-1}\nabla_{\!r}^2
(r\rho) \;\;\mbox{and}\;\;\kappa_{\!_R}
= \kappa_{r}^{} \delta (r-d),
\end{equation}
where $ D_{\!p\,} = D_{{X_a}}^{} \! + D_{{X_b}}^{}, \; ( D_{\!p\,}
\sim \lambda^2 w_h),$ is the coefficient of relative diffusion of
particles $X_a$ and $X_b$, $\:\nabla_{\!r} =
\partial/\partial r,$ and $ \hat \kappa_{\!_R}$ is defined by
$\hat \kappa_{\!_R} = \frac{1}{2}\kappa_r^{} ( {P}_{\!_R}^{}\rho +
\rho{P}_{\!_R}^{})$ [eq. (\ref{gen3})]. In the free diffusion model
the SLE should be solved with the reflective inner boundary
condition at the distance of closest approach $d$:\cite{Shu1}
$\nabla_{\!r}^{}\rho (r,t)|_{r=d} = 0$, and the outer condition
$\rho (r \to \infty,t) \to 0$.

The ESP of $D$-particle is expressed by formula\cite{Shu1,Adr}
\begin{equation} \label{gen8a}
{\cal P} =  2 \Big|\int_{r>d} \! d^{3\!}r\,{\rm Tr}
[S_{_{\!D_{\!z}}} \rho(r, t \to \infty)]\Big|
\end{equation}

The kinetics of ESP generation in the free diffusion model has
already been analyzed in detail for both $D\!-\!D$ and $T\!-\!D$
processes\cite{Goud,Shu1,Adr,Shu2,Shu3} that is why in what follows
we will restrict ourselves to brief summary of some results of this
analysis interesting for our work.

\section{$D\!-\!D$ recombination}

The $D\!-\!D$-recombination is described by the kinetic scheme
(\ref{gen8}) with $X_{a} = D_a$ and $X_b = D_b$, in which $\hat
K_{\!_R}$ is the operator of reaction in the singlet ($S$) state of
the $D_a\!-\!D_b$ pair, given by eq. (\ref{gen3}) with $P_{\!_R}^{}
= P_{\!_S}^{} = |S \rangle \langle S|$.

The most important mechanism of ESP generation in this type of
reactions is known as $ST_{-}$-mechanism.\cite{Clos,Kap} Here we
will consider the specific features of this mechanism in organic
semiconductors.

\subsection{Interactions}

The spin Hamiltonian of the $D_a\!-\!D_b$ pair is assumed to be
represented as a sum
\begin{equation} \label{ehr1}
H (r,t) = H_0 (r) + V(t),
\end{equation}
where $H_0 (r)$ and $V(t)$ are the steady state and randomly
distributed (and may be fluctuating) parts, respectively.

The steady state part of the Hamiltonian
\begin{equation} \label{ehr2}
H_0 (r) = H_{z} + H_{ex} (r)
\end{equation}
is diagonal in the basis of the total electron spin $\hat {\bf S} =
\hat {\bf S}_a + \hat {\bf S}_b$ (called $ST$-basis) and consists of
the Zeeman interaction (in the external magnetic field ${\bf B}$)
\begin{equation} \label{ehr3}
H_{z} = \Omega_0 (S_{a_z}+S_{b_z}),
\end{equation}
in which $\Omega_0 = \bar g \beta B$ is the Zeeman splitting (we
assume that the isotropic parts of $g$-factors of both particles are
the same: $\bar g_a = \bar g_b = \bar g = 2$), and the exchange
interaction
\begin{equation} \label{ehr4}
H_{ex} = J(r)
(\mbox{$\frac{1}{2}$} + 2 {\bf S}_{a}{\bf S}_{b})
\end{equation}
exponentially decreasing with $r$:
\begin{equation} \label{ehr5}
J(r) = J_0 \exp [-\alpha (r-d)],\; (J_0 > 0),
\end{equation}
where $d$ is the distance of closest approach: $d \gg \alpha^{-1}$.

In the $ST_{-}$-mechanism the ESP is assumed to result from quantum
transitions between $U_{\!_S} (r)$ and $U_{_{T_{\pm}}}(r)$ terms of
$\hat H_0^{} (r)$ near the $ST_{\!_{-}}$-crossing region at $r =
r_t^{}$ (see Fig. 1). These transitions are caused by the
fluctuating part $V (t)$ of the Hamiltonian  (non-diagonal in the
$ST$-basis), which is a sum of two terms:
\begin{equation} \label{ehr6}
V(t) = V_{\!_H}(t) + V_{\!_Z} (t).
\end{equation}

\begin{figure}
\setlength{\unitlength}{1cm}
\includegraphics[height=5.0cm,width=7.5cm]{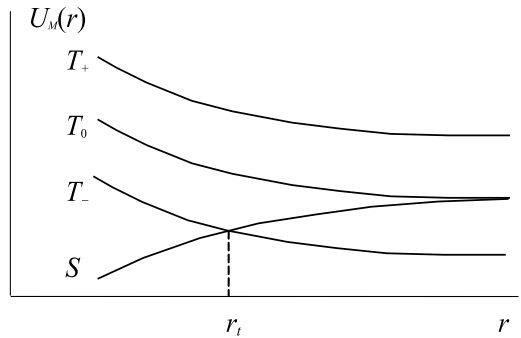}
\caption{The schematic picture of terms $U_{\!_M}^{}(r), \: (M = S,
T_{\!_0}^{}, T_{\!_\pm})$, of the spin Hamiltonian $H_0^{}(r)$ [eq.
(\ref{ehr2})] of the $D\!-\!D$ pair (for $J_0^{}
> 0$), in which $r_t^{}$ is the coordinate of the region of
efficient quantum transitions.}
\end{figure}

The first term $V_{\!_H}(t)$ represents the hyperfine interaction
(HFI), which in the considered realistic case of a large number of
nuclei can quite accurately be treated as the interaction of
electron spins ${\bf S}_a$ and ${\bf S}_b$ with (classical) random
magnetic fields ${\bf B}_a$ and ${\bf B}_b$, respectively:
\begin{equation} \label{ehr7}
V_{\!_H} (t) = \bar g \beta [{\bf S}_{a}{\bf B}_{a}(t)
+ {\bf S}_{b}{\bf B}_{b}(t)].
\end{equation}
Distributions of fields are ${\bf B}_a$ and ${\bf B}_b$ isotropic,
and Gaussian with $\langle {\bf B}_{\nu} \rangle = 0$ and mean
squares $\langle {B}_{\nu}^2 \rangle \sim \sum_{j}
I_{\nu_{j}}(1+I_{\nu_{j}}) a_{\nu_{j}}^2,\: (\nu = a, b)$,
determined by hyperfine coupling constants $a_{\nu_{j}} $.\cite{Wol}
Hopping of $D_a$ and $D_b$ particles results in sudden changing of
nuclear magnetic fields, which can be considered as stochastic
vectors ${\bf B}_{\nu}  (t), \; (\nu = a, b)$. For simplicity, the
correlation functions of projections ${B}_{\nu_{q}} (t), \: (q =
x,y,z)$ are taken in the form $\langle {B}_{\nu_{q}}(t)
{B}_{\nu_{q'}}(0)\rangle \sim \delta_{qq'}\langle {B}_{\nu}^{2}
\rangle \exp (-w_{\!_{H_\nu}} t)$ (so that $\langle
V_{\!_{H}}(t)V_{\!_{H}} (0)\rangle \sim e^{-w_{\!_{H_{\nu}}}\! t}$).

The second term $ V_{\!_Z}(t) $ in eq. (\ref{ehr6}) is the
anisotropic part of the Zeeman interaction (AZI):
\begin{equation} \label{ehr8}
V_{\!_Z} (t) = \beta [{\bf S}_{a} \hat g'_{a}(t) {\bf B} +
{\bf S}_{b} \hat g'_{b}(t) {\bf B}],
\end{equation}
where
\begin{equation} \label{ehr9}
\hat g'_{\nu} = \hat g_{\nu} - \bar g_{\nu}\;\; \mbox{with} \;\;
\bar g_{\nu}  = {\rm Tr} (\hat g_{\nu}), \;\;(\nu = a, b).
\end{equation}
The representation (\ref{ehr8}) implies that ${\rm Tr} (\hat
g'_{\nu}) = 0$. As in the case of the HFI the time dependence $\hat
g'_{\nu} (t)$ results from $D$-particle hopping. We will use the
simple form of the correlation function\cite{Car1} $\langle
g'_{\nu_{qq'}}(t) g'_{\nu_{q'q}} (0)\rangle \sim
e^{-w_{\!_{Z_{\nu}}}\! t},\; (q \neq q')$, (which means that
$\langle V_{\!_{Z}}(t)V_{\!_{Z}} (0)\rangle \sim
e^{-w_{\!_{H_{\nu}}}\! t}$).

\subsection{ESP generation kinetics}

In order to evaluate the ESP ${\cal P}$ we need to specify the
initial spin state $\rho_0^{}$ [see eq. (\ref{gen4a})]. In our
analysis, for certainty, we consider the equilibrium (isotropic)
state:
\begin{equation} \label{ehr9a}
\rho_0^{} = \mbox{$\frac{1}{4}$} (P_{\!_S} + P_{\!_T}),
\end{equation}
where $P_{_{\!_S}} = |S \rangle \langle S|$ and $P_{\!_T} =
\sum_{_{m =0, \pm}} |T_{\!_m}\rangle \langle T_{\!_m}|$. Note, that
for other initial spin states the ESP is, actually, proportional to
that for $\rho_{0}^{}$ (\ref{ehr9a}).\cite{Shu6}

\subsubsection{Cage model}

We have already noted that quantum transitions, resulting in the
ESP, are localized in the narrow region of width $\delta r \sim
\alpha^{-1} \ll d$ (recall that $d \sim \lambda$). In such a case
the ESP generation can properly be described by treating spin
evolution of $D\!-\!D$ pairs in all positions of them within this
region as cages, whose dissociation rate is associated with the
hopping rate $w_h^{}$ (see Sec. II): $w_d^{} \lesssim w_h^{}$.

The interaction $V$ in thus introduced cages should be considered as
time independent though randomly distributed, since the fluctuations
of $V(t)$ are formed just as a result of hopping motion, i.e. at
times $t \gtrsim \tau_h^{} = w_h^{-1}$.

Under the assumption of homogeneous distribution of $D\!-\!D$
coordinate ${\bf r}$ in the cage (see Sec. II) and with the use of
eq. (\ref{gen7b}) one obtains the following formula for the ESP
${\cal P}$, generated in $D\!-\!D$ recombination,
\begin{equation} \label{cage0a}
{\cal P} = |{\cal P}_{\!_-}  -  {\cal P_{\!_+}}| =
|\langle p_{_{-+}}(r) \rangle_{_{{\bf r}\in {\cal V}_c^{}}}\!|,
\end{equation}
where ${\cal P}_{\!_\pm} = w_d^{}\langle\widetilde{\rho}_{\pm}^{}
(r)\rangle_{_{{\bf r}\in {\cal V}_c^{}}}$ are the contributions of
$T_{\pm}\!\! \leftrightarrow \! S$ transitions, in which
$\widetilde{\rho}_{\pm}^{}(r) = \langle T_{_\pm}|\widetilde{\rho}
(r,\epsilon = 0) |T_{_\pm}\rangle$, and $p_{_{-+}}(r) = w_d^{}
[\widetilde{\rho}_{-}^{} (r) - \widetilde{\rho}_{+}^{} (r)]$.

Solution of the SLE in the second order in the randomly distributed
interaction $V$ yields formula
\begin{equation} \label{cage0}
p_{_{-+}} (r) = \bar p \,\langle |V_{\!_{\pm}}|^2\rangle\!
\bigg[\frac{1}{w_c^2 + \Omega_{\!_-}^{2}(r)}
- \frac{1}{w_c^2 + \Omega_{\!_+}^{2}(r)}\bigg]
\end{equation}
with $|V_{\!_{\pm}}| = |\langle S| V |T_{_+}\rangle| = |\langle S| V
|T_{_-}\rangle|, \;\Omega_{\!_\pm}(r) = \Omega_0 \pm 2J(r)$,
$\,w_{c} = w_d + \frac{1}{2}\kappa_{\!_R}$ being the
$S\!-\!T$-dephasing rate in the cage, and $\bar p =
\frac{1}{2}(\kappa_{\!_R}w_c^{})/[w_d(w_d + \kappa_{\!_R})]$.

Substituting this formula into eq. (\ref{cage0a}) one can obtain the
following expression for the SPE ${\cal P}$, corresponding to the
initial equilibrium spin state (\ref{ehr9a}):
\begin{equation} \label{cage1}
{\cal P} = p_{c}^{} |\varphi_{\omega_{\!_-}}^{}\!\!
- \varphi_{\omega_{\!_+}}^{}\! |
[{\omega}_{c}^{}/(1 + {\omega}_{c}^2)],
\end{equation}
in which $\:\varphi_{\omega_{\!_\pm}}^{} = {\rm
arctan}(\omega_{\pm}^{})+ (2 {\omega}_{c}^{})^{-1}\ln (1 +
\omega_{\pm}^2)\,$ with $\, \omega_{_\pm} = (\Omega_0^{} \pm
2J_0^{})/w_c^{}$,
\begin{equation} \label{cage2}
{\omega}_{c}^{} = \Omega_0^{} / w_{c}^{} \;\,\mbox{and}\;\,
p_{c}^{} \approx \mbox{$\frac{1}{2}$} p_{\!_{R_c}}^{}
\!\bigg(\!\frac{\langle |V_\pm^{}|^2\rangle}{w_{c} w_d}\!\bigg)\!
\bigg(\!\frac{4\pi d^2}{\alpha  {\cal V}_c^{}}\bigg).
\end{equation}
Here  $p_{\!_{Rc}} = \kappa_{\!_R}/(w_d + \kappa_{\!_R})$ is the
probability of reaction in $S$-state and ${\cal V}_c^{}$ is the
volume of the cage.

Note that in the presence of $V(t)$-fluctuations the rate formulas
(\ref{cage1}) and (\ref{cage2}) are still valid but $\,w_{c} = w_d +
\frac{1}{2}\kappa_{\!_R}  + w_{\mu}^{}$, $(\: \mu = H, Z)$, where
$w_{\mu}$ is the rate of decay of correlation function $\langle
V_{\mu}^{}(t)V_{\mu}^{}(0) \rangle$, introduced above.

It also is worth noting that the ESP for any other initial spin
state is closely related to that (\ref{cage1}) for the equilibrium
state $\rho_{_E}$ (\ref{ehr9a}). For example, for the triplet
initial state $\rho_0^{} = \rho_{_T} = \frac{1}{3} P_{_T}$ the ESP
${\cal P}(\rho_{_T}) = (4/3){\cal P}(\rho_{_E})/p_{\!_{R_c}}^{}$.

Formula (\ref{cage1}) is fairly cumbersome. Fortunately, in
applications it is quite sufficient to use simpler limiting
expressions at large $J_0 > \Omega_0, w_c^{}$ and small $J_0 <
\Omega_0, w_c^{}$:
\begin{eqnarray}
{\cal P}={\cal P}_{\!s}^{}&=& (\pi p_{c}^{})
\frac{{\omega}_{c}^{}}{1 + {\omega}_{c}^2}
\,\qquad\quad\;\;\,\mbox{for} \;
J_0 > \Omega_0, w_c^{}, \quad\; \label{cage4a}    \\
{\cal P}={\cal P}_{\!w}^{}&=& 8 p_{c}^{}
\bigg(\frac{J_0^{}}{w_c^{}}\bigg)
\frac{{\omega}_{c}^{}}{(1 + {\omega}_{c}^2)^2} \;\;\,
\mbox{for} \;J_0 < \Omega_0,  w_c^{}.\;\;\;\;\label{cage4b}
\end{eqnarray}

It is easily seen that the considered two cases $J_0 > \Omega_0$ and
$J_0 < \Omega_0$ correspond $r_t^{} > d$ and $r_t^{} < d$,
respectively, where $r_t^{}$ is the distance of $ST_{\!_-}$ crossing
(see Fig. 1).

\subsubsection{Free diffusion model}

Simple formulas for the ESP have also been derived in the free
diffusion model.\cite{Goud,Shu1,Adr,Shu2,Shu3} Naturally, the ESP,
generated in diffusion controlled $D\!-\!D$ recombination, depends
(but weakly) on the initial distance $r_i^{}$. Below, for certainty,
we present the formulas in the realistic case $r_i^{} \approx d$.

The most interesting expressions for the ESP at large $J_0^{}
> \Omega_0^{}, w_{f}^{}$ and small $J_0^{} <\Omega_0^{}, w_{f}^{}$
[$w_{f}^{}$ is the correlation rate of $V(t)$ fluctuations (see
below)] turn out to be similar to those obtained above in the cage
model:\cite{Goud,Shu1,Adr,Shu2,Shu3}
\begin{eqnarray}
{\cal P}={\cal P}_{\!s}^{}&=& (\pi p_{f}^{})
\frac{{\omega}_{\!f}^{}}{1 + {\omega}_{\!f}^2} \,\qquad\quad\:\,\mbox{for} \;
J_0 > \Omega_0, w_{\!f}^{},
\quad\;\; \label{dif1a}\\
{\cal P}={\cal P}_{\!w}^{}&=& 8 p_{f}^{}\!
\bigg(\frac{J_0^{}}{w_{\!f}^{}}\bigg)\!
\frac{{\omega}_{\!f}^{}}{(1 + {\omega}_{\!f}^2)^2} \;\;\,
\mbox{for} \;J_0 < \Omega_0^{}, w_{\!f}^{}.
\quad\;\;\;\label{dif1b}
\end{eqnarray}
Here
\begin{equation} \label{dif2}
{\omega}_{\!f}^{} = \Omega_0^{}/w_{\!f}^{} \;\:\mbox{and}\;\:
p_{f}^{} \approx \mbox{$\frac{1}{2}$}p_{_{\!R_f}}^{}
\bigg(\frac{\langle |V_\pm|^2\rangle}{w_{\!f}^2}\bigg)
\bigg(\frac{d w_{\!f}^{}}{\alpha D_{\!p}}\bigg)
\end{equation}
with $w_f = w_{\mu}$, $(\: \mu = H, Z)$, being the decay rate of the
correlation function $\langle V_{\mu}^{}(t)V_{\mu}^{}(0) \rangle$
(defined in Sec. IIIA).


Similarly to the cage model $p_{_{\!R_f}}^{} = l_{_S}/r_i^{} \approx
l_{_S}/d$ is the probability of (diffusion controlled) reaction in
$S$-state with the rate $\kappa_{\!_R}$, given by eq. (\ref{gen8}),
in which $l_{_S} = d(\kappa_{r}d/D_{\!p})[1 +
(\kappa_{r}d/D_{\!p})]^{-1}$ is the reaction radius.

\subsection{Some general properties of $D\!-\!D$ ESP}

The important specific features of the ESP for different limits and
mechanisms (discussed above) are conveniently represented in terms
of the dimensionless functions $\Phi_{\mu}^{\chi} (\omega)$, in
which the superscript $\mu$ specifies the ESP mechanism (see Sec.
IIIA): HFI ($\mu = H$) and AZI ($\mu = Z$), while the subscript
$\chi$ indicates the value of $J_0$: large ($\chi = s$) and small
($\chi = w$). In accordance with above-obtained expressions, one can
write
\begin{equation} \label{pro1}
{\cal P} = p_{\!_{R_{\!\gamma}}}^{}\langle v_{\!\mu}^2\rangle
\bar P_{\!\chi} \Phi_{\mu}^{\chi} (\omega_\gamma^{}\!).
\end{equation}
Here  $p_{\!_{R_{\!\gamma}}}^{}$ and $\omega_\gamma$ are defined in
eqs. (\ref{cage2}) and (\ref{dif2}) for the cage ($\gamma = c$) and
free diffusion ($\gamma = f$) models, respectively. The parameters
$\langle v_{\!\mu}^2\rangle$ are the dimensionless squares of
interactions defined by formulas
\begin{equation} \label{pro1a}
\langle v_{\!_H}^2\rangle = (\bar g\beta)^2 \frac{\langle B_a^2\rangle \!
+ \!\langle B_b^2 \rangle}{12 w_{\gamma}^{2}},\;
\langle v_{\!_Z}^2\rangle =
\frac{g'_a\!:\!g_a'\!+\!g'_b\!:\!g_b'}{40 \bar g},
\end{equation}
with $w_{\gamma}^{}, \,(\gamma = c, f),$ given in eqs. (\ref{cage2})
and (\ref{dif2}).

{\it a. Strong exchange interaction limit $(\chi = s)$.}  In the
limit of strong exchange interactions (large $J_0^{}$) we get for
the HFI ($\mu = H$) and AZI ($\mu = Z$) mechanisms
\begin{equation} \label{pro2}
\Phi_{\!_H}^{s} (\omega) = \omega/(1 + \omega^2), \;\;\;
\Phi_{\!_Z}^{s} (\omega) = \omega^3/(1 + \omega^2),
\end{equation}
where the parameter $\bar P_{s}$ is written for the cage ($\gamma =
c$) and free diffusion ($\gamma = f$) models as follows:
\begin{eqnarray}
\bar P_{s}^{}  &=& (\mbox{$\frac{1}{2}$}\pi)
(w_{c}/w_d) (4\pi d^2/\alpha  {\cal V}_c^{})
\;\;\;\,\mbox{for}\;\; \gamma = c, \qquad\label{pro3a}\\
\bar P_{s}^{}  &=& (\mbox{$\frac{1}{2}$}\pi)
(d w_{\!f}^{}/\alpha D_{\!p}^{})
\;\;\:\,\qquad\qquad\mbox{for} \;\; \gamma = f. \:\label{pro3b}
\end{eqnarray}

{\it b. \,Weak exchange interaction limit $(\chi = w)$.}  In the
weak exchange interaction limit (small $J_0^{}$) one obtains
\begin{equation} \label{pro6a}
\Phi_{\!_H}^{w} (\omega) = \omega/(1 + \omega^2)^2, \;\;\;
\Phi_{\!_Z}^{w} (\omega) = \omega^3/(1 + \omega^2)^2,
\end{equation}
with
\begin{equation} \label{pro6b}
\bar P_{\!w}  = (8/\pi) (J_0/w_{\gamma})\bar P_{\!s},
\end{equation}
both in the cage and free diffusion models (for $\gamma = c, f$).

The functions $\Phi_{\mu}^{\chi} (\omega)$ are shown in Fig. 2.
Notice that all dependences $\Phi_{\mu}^{\chi} (\omega)$, except
$\Phi_{\!_Z}^{s} (\omega)$,  are non-monotonic with maximum at
$\omega \sim 1$.

\begin{figure}
\setlength{\unitlength}{1cm}
\includegraphics[height=8.3cm,width=6.0cm]{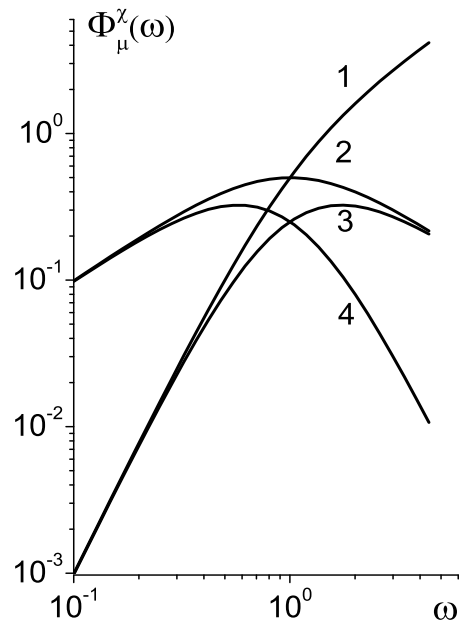}
\caption{The functions $\Phi_{\mu}^{\chi} (\omega)$, which determine
the ESP dependence on the Zeeman splitting $\omega =
\Omega_0/w_{\gamma}$ in cage ($\gamma = c$) and free diffusion
($\gamma = c$) models [see eqs. (\ref{pro2}) and (\ref{pro6a})]: (1)
$\Phi_{\!_Z}^{s} (\omega)$, (2) $\Phi_{\!_H}^{s} (\omega)$, (3)
$\Phi_{\!_Z}^{w} (\omega)$, and (4) $\Phi_{\!_H}^{w} (\omega)$ [see
eqs. (\ref{pro2}) and (\ref{pro6a})]. In these functions the
superscript $\chi$ and subscript $\mu$ specify the exchange
interaction strength and the mechanism of the ESP generation,
respectively, ($\chi = s, w$ and $\mu = H, Z$).}
\end{figure}

Note that in the free diffusion model the formulas for the ESP are
derived under the assumption of not very high mobilities of
particles (fairly small diffusion coefficients) so that
$\xi_{\Omega}^{} = \Omega_0 d^2/D_{\!p}^{}
> 1$.\cite{Shu1} In the opposite limit $\xi_{\Omega}^{} = \Omega_0
d^2/D_{\!p}^{} < 1$ the ESP is essentially determined by the
parameter $\xi_w^{} = w_c^{}d^2/D_{\!p}^{}$. For $\xi_w^{}
> 1$ we get the results presented above independently of the value
of $\xi_{\Omega}$. For $\xi_w^{} \ll  1$, however, the behavior of
$\Phi_{\mu}^{\chi} (\omega)$ at $\xi_{\omega} < 1$ becomes
non-analytic: for example, $\Phi_{\!_H}^{s} (\omega) \sim
\omega^{-3/2}$.\cite{Shu1}

\section{$T\!-\!D$ quenching}

The $T\!-\!D$ quenching is another example of spin selective
processes, leading to the generation of the ESP of $D$-particles.
The kinetic scheme of the geminate $T\!-\!D$ process can be
represented in the form (\ref{gen0}) with $X_a = T$ and $X_b = D$
(in OSCs $D$-particle is associated with trapped electron).

The quenching is believed\cite{Ern} to result in the transition to
$X_{_R} = (S_0^{}D^{*})_{D}$, where $S_0^{}$ is the molecule in the
ground (singlet) state, and $D^{*}$ is the excited doublet (free
electron), i.e. the quenching rate is expected to be non-zero only
in the doublet ($|D_{\!_{\pm1/2}} \rangle$) states of $T\!-\!D$
pair, corresponding to the total spin $S = 1/2$. In this case the
quenching operator $\hat K_{\!_R}$ is represented in the form
(\ref{gen3}) with $P_{\!_R}^{} = P_{\!_D}^{} = |D_{\!_{+1/2}}
\rangle \langle D_{\!_{+1/2}}| + |D_{\!_{-1/2}} \rangle \langle
D_{\!_{-1/2}}|$.

It is clear from the above comments to the general scheme
(\ref{gen0}) that, in principle, there are two types of
$D$-particles involved in the process under study: $D$ and $D^{*}$
[escaped from the reaction ($D$) and resulting from reaction
($D^{*}$)]. In OSCs these particles are associated with trapped
electrons and electrons in in the conducting band,
respectively.\cite{Ern} The $T\!\!-\!\!D$ quenching gives rise to
the ESP of both particles. In the analysis, however, we will assume
that the kinetic and MFE parameters of these types of $D$-particles
are different, i.e. the contributions of them to the ESP are
experimentally distinguishable.

In our work we will concentrate on the evaluation of the ESP of
$D^{*}$ particles.

The methods of estimating ESP of both $D$ particles for liquid phase
processes, i.e. in the diffusion model of relative motion of
particles, have already been discussed in a number of
papers.\cite{Goud,Shu2,Shu3} Below we will analyze some specific
features of the ESP generation as applied to disordered
semiconductors, considering more thoroughly the cage model, as an
example.

\subsection{Interactions}

In our analysis we assume that the spin Hamiltonian of the $T\!-\!D$
pair (similar to that for  $D\!-\!D$ systems) can be represented as
a sum $H (r,t) = H_0 (r) + V(t),$ where $H_0 (r)$ and $V(t)$ are the
steady state and randomly distributed (and may be fluctuating)
parts, respectively.

The steady state part
\begin{equation} \label{td1}
H_0 (r) = H_{z} + H_{ex} (r),
\end{equation}
diagonal in the basis of the total electron spin ${\bf S} = {\bf
S}_{_T} + {\bf S}_{_D}$, is a sum of the Zeeman and exchange
interactions:
\begin{equation} \label{td2}
H_{z} = \omega_0 S_z^{} \;\; \mbox{and} \;\;
H_{ex} = J(r)
\big(\mbox{$\frac{1}{2}$} + 2 {\bf S}_{_T}{\bf S}_{_D}\big),
\end{equation}
in which $S_z^{} = S_{_{T_z}}+S_{_{D_z}},\,$ $\omega_0 = g \beta B$
(we assume that $g_{_D} = g_{_T} = g = 2$), and $J(r)$ is given by
eq. (\ref{ehr5}). The terms $U_{\!_M} (r), \: (M = Q_{\!_m},
D_{\!_m})$, of the Hamiltonian $H_0^{}(r)$ are displayed in Fig. 3.

Transitions in the regions of crossing of $U_{\!_M} (r)$-terms are
determined by the fluctuating zero-field-splitting (ZFS) interaction
\begin{equation} \label{td3}
V = V_{\!_T} = D_{_T}(S_{_{T_z'}}^2 -
\mbox{$\frac{1}{3}$}S_{_{\! T}}^{2})
+ E_{_T} (S_{_{T_x'}}^2 - S_{_{T_y'}}^2),
\end{equation}
where $S_{_{T_j}}, \, (j = x',y',z'),$ are the projections of the
spin of $T$-exciton on the eigenaxes of the ZFS tensor. Usually
$D_{_T} \gg E_{_T}$, that is why below we will neglect the terms
$\sim E_{_T}$, taking $V_{_{T}} \approx D_{_T}S_{_{T_z'}}^2 $. The
fluctuations result from hopping motion of $T$ exciton over sites
with randomly oriented directions of axes $x',y',z'$. These
fluctuations show themselves in those of interaction matrix elements
$V_{\!_{T_{MM'}}} (t) \equiv \langle M |V_{\!_T} (t)| M'\rangle$,
correlation function of which is assumed to be of the
form\cite{Car1} $\langle V_{\!_{T_{MM'}}}^{*} \!(t)
V_{\!_{T_{MM'}}}\! (0) \rangle_{\!_{\Omega'}} = \langle
|V_{\!_{_{MM'}}}^2| \rangle e^{-w_{\!_T} t}$ (here $M, M' =
Q_{\!_m}, D_{\!_m}$ and $\langle \dots \rangle_{\!_{\Omega'}} $
denotes averaging over orientations of the ZFS eigenaxes).
\begin{figure}
\setlength{\unitlength}{1cm}
\includegraphics[height=6.2cm,width=7.0cm]{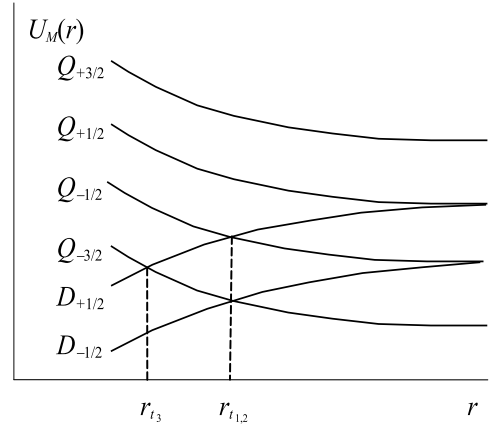}
\caption{The schematic picture of terms $U_{_M}^{}(r), \: (M =
D_{\!_{\pm 1/2}}, Q_{\!_{\pm 1/2}}^{}, Q_{\!_{\pm 3/2}})$, of the
spin Hamiltonian $H_0^{}(r)$ [eq. (\ref{td1})] of the $T\!-\!D$ pair
(for $J_0^{} > 0$), in which $r_{t_{1,2}}^{}$ and $r_{t_3}^{}$ are
the coordinates of the regions of efficient quantum transitions.}
\end{figure}

\subsection{ESP generation kinetics}

In $T\!-\!D$ system the ESP is assumed to result from quantum
transitions in the regions of crossing of terms $U_{Q_{_{\!\pm
3/2}}}(r)$ and $U_{Q_{_{\!\pm 1/2}}}(r)$ with reactive terms
$U_{D_{_{\!\pm 1/2}}}(r)$. These regions are indicated in Fig. 3.

Similarly to the $D\!-\!D$ ESP, the ESP in $T\!-\!D$ quenching will
be evaluated for the equilibrium initial spin state
\begin{equation} \label{td4a}
\rho_0^{} = \rho_{_E} = \mbox{$\frac{1}{6}$} (P_{\!_D} + P_{\!_Q}),
\end{equation}
where $P_{\!_D}^{} $ and $P_{\!_Q}^{} $ are the operators of
projection onto the subspaces of $D$- and $Q$-states, respectively:
$P_{\!_D}^{} = \sum_{_{m = \pm 1/2}} |D_{\!_{m}} \rangle \langle
D_{\!_{m}}|$ and $P_{\!_Q} = \sum_{_{m =\pm 1/2, \pm 3/2}}
|Q_{\!_m}\rangle \langle Q_{\!_m}|$. The ESP for other initial spin
state are, actually, proportional to that for
$\rho_{_E}$.\cite{Shu6}

In the majority of $T\!-\!D$ processes the above-mentioned
ZFS-induced quantum transitions can be calculated perturbatively in
the second order in $V$.  In this approximation the effect of
transitions is represented as a sum of contributions of all pairs of
coupled terms, which can be evaluated with formulas derived above
for the $D\!-\!D$ ESP ${\cal P}$.\cite{Shu1} It is worth noting that
these formulas represent the effect of quantum transitions on the
$D\!-\!D$ ESP of escaped articles. However, similar expressions are,
certainly, valid for the product particles as well.\cite{Shu3}

Analysis of quantum transitions between terms (displayed in Fig. 3)
shows that the contributions to the ESP ${\cal P}_{\!*}^{}$ of the
$D^{*}$-particle can be represented as a sum of three contributions
of type of (\ref{cage0}):
\begin{equation} \label{td4}
{\cal P} = {\cal P}_{\!*}^{} = \mbox{$\frac{2}{3}$}
\big|\delta{\cal P}_{_{1/2,3/2}} + \delta{\cal P}_{_{1/2,1/2}} +
\delta{\cal P}_{_{1/2,-3/2}}\big|\,,
\end{equation}
in which $\delta{\cal P}_{\!_{\mu,\mu'}}$ are defined by
\begin{equation} \label{td5}
\delta{\cal P}_{\!_{m,m'}} =
{\cal P}_{\!_{D_{\!-m^{}}Q_{-m'}}}\!\!\! - \,{\cal P}_{\!_{D_m^{}Q_{m'}}},
\end{equation}
[with $m = 1/2, \,m' = 1/2, \, 3/2$], where ${\cal P}_{\!_{D_m^{}
Q_{m'}}}$ is the contribution of the transition $D_{\!_m^{}}
\leftarrow Q_{\!_{m'}}$.

\subsubsection{Evaluation of ESP in the cage model}

In the cage model we can obtain the expression for $\delta{\cal
P}_{\!_{m,m'}}$ in terms of the distance dependent solution
$p_{\!_{mm'}}^{} (r)$ similar to that derived above [see eq.
(\ref{cage0})]
\begin{equation} \label{td6a}
\delta{\cal P}_{\!_{m,m'}} =
\langle p_{_{\!mm'}} (r) \rangle_{_{{\bf r}\in {\cal V}_c^{}}}.
\end{equation}
In this formula
\begin{equation} \label{td6}
p_{\!_{mm'}}^{} (r) = \bar p\,\langle |V_{\!_{mm'}}^2| \rangle\!
\bigg[\frac{1}{w_c^2 + \Omega_{\!-}^{2}\!(r)}
- \frac{1}{w_c^2 + \Omega_{\!+}^{2}\!(r)}\bigg],
\end{equation}
where the parameter $\bar p = \frac{1}{2}(\kappa_{\!_R}
w_c)/[w_d^{}(w_d^{} + \kappa_{\!_R})]$ is identical to that in eq.
(\ref{cage0}), and
\begin{equation} \label{td7}
\Omega_{\!\pm}^{}(r) = \alpha_{\!_{mm'}}^{}\Omega_0^{} \pm 2J(r)
\end{equation}
with $\alpha_{\!_{\pm 1/2,\pm 3/2}} = \alpha_{\!_{\pm 1/2,\mp 1/2}}
= 1$ and $\alpha_{\!_{\pm 1/2,\mp 3/2}}  = 2$. As for $\langle
|V_{\!_{mm'}}^2|\rangle$, for $m$ and $m'$, corresponding to the
contributions (\ref{td5}), they are written as $\langle |V_{\!_{\pm
1/2,\pm3/2}}^2|\rangle = \frac{1}{45}D_{\!_T}^2, \: \langle
|V_{\!_{\pm 1/2,\mp 1/2}}^2|\rangle = \frac{1}{15}D_{\!_T}^2, \:$
and $\,\langle |V_{\!_{\pm 1/2,\mp 3/2}}^2|\rangle
=\frac{4}{45}D_{\!_T}^2$.

Note that similar to the $D\!-\!D$ process in the presence of
$V_{\!_T}$-fluctuations the dephasing rate is written as $\,w_{c} =
w_d + \frac{1}{2}\kappa_{\!_R} + w_{\!_T}^{}$ [rather than as that
defined in eq. (\ref{pro1})], where $w_{\!_T}^{}$ is the decay rate
of the correlation function $\langle V_{\!_{T_{MM'}}}^{*} \!(t)
V_{\!_{T_{MM'}}} \!(0) \rangle_{\!_{\Omega'}}$, defined in Sec IVA.

Substitution of all terms (\ref{td6}) into eq. (\ref{td4}) yields
the final expression for the ESP of $D^{*}$ (see below).

\subsubsection{Combined formula for ESP, predicted by two models}

The $T\!-\!D$ ESP, predicted in the above-analyzed cage model and
previously studied free diffusion model,\cite{Goud,Shu2,Shu3} can be
represented in one combined expression similar to that (\ref{pro1})
obtained for the $D\!-\!D$ ESP
\begin{equation} \label{td9}
{\cal P}_{\!*}^{} = \mbox{$\frac{2}{3}$}p_{\!_{R\!_\gamma}}^{}\!
\langle v_{\!_T}^2\rangle
\bar P_{\!\chi} [\Phi_{\!_H}^{\chi}(\omega_{\!\gamma}^{}) +
2\Phi_{\!_H}^{\chi}(2\omega_{\!\gamma}^{})], \;(\chi = s,w).
\end{equation}
In this expression $\langle v_{\!_T}^2\rangle =
\mbox{$\frac{2}{45}$} (D_{\!_T}^2/w_{c}^2)$, the functions
$\Phi_{\!_H}^{\chi}(\omega_{\!\gamma}^{})$ are defined in eqs.
(\ref{pro2}) and (\ref{pro6a}), and the parameters $\bar
P_{\!\chi}^{}$ are given in eqs. (\ref{pro3a}), (\ref{pro3b}), and
(\ref{pro6b}).

The difference between the cage and free diffusion ESP lies in the
parameters $p_{\!_{R\!_\gamma}}^{}$ and $\omega_{\gamma}^{}$: for
the cage ESP ($\gamma = c$) $p_{\!_{R\!_c}}^{} =
\kappa_{\!_R}/(w_d^{} + \kappa_{\!_R})$ and $\omega_{c}^{} =
\Omega_0^{}/w_{c}^{}\,$ [see eq. (\ref{cage2})], while for free
diffusion ESP ($\gamma = f$) $p_{\!_{R\!\!_f}}^{} =
(\kappa_{\!_r}d/D_p^{})/[1 + (\kappa_{\!_r}d/D_p^{})]$ and
$\omega_{\!f}^{} = \Omega_0^{}/w_{\!f}^{}\,$ with $w_{\!f}^{} =
w_{\!_{T}}$ [eq. (\ref{dif2})].

\medskip

It is worth noting that the dependence of ${\cal P}_{\!*}^{}$ on
$\omega_{\gamma}^{},$ (i.e. on $\Omega_0^{}$), predicted by both the
cage and free diffusion models, is determined by that of
$\Phi_{\!_H}^{\chi} (\omega)$, i.e. function ${\cal P}_{\!*}^{}
(\omega_{\gamma}^{})$ is non-monotonic the maximum of ${\cal
P}_{\!*}^{} (\omega_{\gamma}^{})$ is located at $\omega_{\gamma}^{}
= \omega_{\gamma_m}^{} \sim 1$ (see Fig. 2).

\section{Discussion}

The obtained formulas allow us to analyze easily the specific
features of the ESP in $D\!-\!D$ and $T\!-\!D$ processes.

\subsection{General remarks}

Before estimating the ESP some remarks are needed on the validity of
models and approximations applied.

1. So far in our work we have discussed the absolute value ${\cal
P}$ of the ESP. As for the ESP sign, it can be characterized by the
sign factor $S_{\!_{\cal P}} = \pm 1$, defined by the relation $
2\langle S_{_{\!Dz}}^{} \rangle = S_{\!_{\cal P}} {\cal P}$, in
which the average $\langle \dots \rangle$ is made over spin state
and distances (see Sec. II). The factor $S_{\!_{\cal P}}$ can easily
be obtained from the qualitative analysis of quantum transitions in
$D\!-\!D$ and $T\!-\!D$ systems. In general, $S_{\!_{\cal P}}$ is
essentially determined by the sign of the exchange interaction
$S_{\!_{\cal P}} \sim {\rm sign} (J_0^{})$. Below we will briefly
discuss the ESP sign in the most realistic case $J_0 > 0$ (for $J_0
> 0$ the terms are shown in Figs.  1 and 3).

a) In the $D\!-\!D$ process the  ESP of survived $D$-particles is
determined by $ST_{-}$-transitions, leading to the decrease of the
population of $T_{-}$-state, i.e. to $S_{\!_{\cal P}} = +1$.

b) In the $T\!-\!D$ quenching the  ESP of $D^{*}$ particles results
from quantum transitions, associated with contributions (\ref{td5})
resulting in larger population of $D_{\!_{+1/2}}$-state (because of
stronger coupling between initial $Q$ states and this $D$-state).
Therefore for $D^{*}$ particles $S_{\!_{\cal P}} = +1$ as well.

2. In the proposed theory the ESP ${\cal P}$ is evaluated in the
second order in $\zeta_{\gamma}^{} = \langle |V^2_{}|
\rangle/w_{\gamma}^{} \ll 1, \:(\gamma = c, f)$, which implies that
${\cal P} \ll 1$. This does not mean, however, that the generated
ESP is small, in general. For example, the effect of attractive
interparticle interactions can lead to strong increase of the ESP
(see Sec. V.B.2).

3. In the cage model the spin selective reaction rate
$\kappa_{\!_R}$ is suggested to be independent of the distance $r$.
This assumption is, in principle, not quite consistent taking into
account that $J(r)$ is treated as distance dependent. Notice, that
the dependence $\kappa_{\!_R}$ on $r$ could be analyzed within
generalized formulas [see eqs. (\ref{cage0}) and (\ref{cage0})],
though, the modification of formulas would result in the strong
complication of the expression for ${\cal P}$. To justify the
simplifying assumption on $\kappa_{\!_R}$ independence of $r$ it is
worth noting that the manifestation of $\kappa_{\!_R}
(r)$-dependence in ${\cal P}$ is fairly weak. It is especially weak
in two above-considered limits of strong and weak exchange
interaction, most important for further applications.

a) For strong exchange interactions the quantum transitions,
determining the ESP, are highly localized in the regions of term
crossing (Figs 1, 3). The distance dependence $\kappa_{\!_R} (r)$ in
these regions can be neglected.

b) In the weak exchange interaction limit the dependence of ${\cal
P}$ on the rate $\kappa_{\!_R}$ results from that of the probability
of reaction in the cage $p_{\!_{R_\gamma}} \sim \kappa_{\!_R},\:
(\gamma = c, f)$. In the free diffusion model the weak effect of the
$r$-dependence of reactivity is demonstrated in earlier
works.\cite{Goud,Shu1,Adr,Shu2,Shu3} Here we will discuss the cage
model, in which this effect shows itself in the average over ${\bf
r}$ in formula (\ref{cage0a}) for ${\cal P}$, i.e. in the average
$\langle p_{\!_{R_c}} \! J \rangle_{\!r}^{},$ where $p_{\!_{R_c}} =
\kappa_{\!_R}/ (w_d^{} + \kappa_{\!_R})$. It is seen that in
$\langle p_{\!_{R_c}} \! J \rangle_{\!r}^{}$ the $\kappa_{\!_R} (r)$
dependence manifests itself only weakly. For example, in the low
reactivity limit, when $p_{\!_{R_c}} \approx \kappa_{\!_R}/ w_d^{}$
and the manifestation is strongest, the dependence $\kappa_{\!_R}(r)
\sim e^{-\alpha r}$, identical to that of $J (r)$ [see eq.
(\ref{ehr5})], results only in two-fold reduction of the ESP. This
effect is too small to be of any importance for our semiquantitative
study of the problem.

\subsection{Estimations of the ESP}

\subsubsection{Bulk processes}

In our analysis we have considered the ESP generation in geminate
processes. It can be known, however, that the geminate ESP ${\cal
P}$, calculated for the equilibrium initial spin state, is directly
related to the rate ${\cal K}$ of ESP generation in the
corresponding bulk processes:\cite{Shu6}
\begin{equation} \label{dis1}
{\cal K} = k_0^{}{\cal P} \;\; \mbox{with}\;\;
k_0^{} = 4\pi r_{\!c}^{} D_{\!p}^{}
\end{equation}
(note that for $T\!-\!D$ system ${\cal P}\equiv {\cal P}_{\!*}^{}$).
In this formula $D_{\!p}^{}$ is the coefficient of relative
diffusion and $r_{\!c}^{}$ is the radius of the cage ($r_c^{} \sim
\lambda$) in the cage model, or the radius of efficient quantum
transitions ($r_{\!c}^{} \sim d$) in the free diffusion model. Our
assumption that $d \sim \lambda$ (Sec. II) means that in both models
the estimations for $r_{\!c}^{}$ are close.

Note that the parameter $r_c^{}$, whose meaning in the diffusion
theory is clear,\cite{Goud,Shu1,Shu2,Shu3} is not quite well defined
in the cage model. This uncertainty of the cage-model definition of
$r_c^{}$ is, however, not very strong and is not essential for our
semiquantitative analysis.

Sometimes it is convenient to relate the rate of ESP generation to
the reaction rate:
\begin{equation} \label{dis1a}
{\cal K}_r^{} \approx (p_s^{} p_{\!_{R\!_\gamma}})k_0^{},
\;\; (\gamma = c, f),
\end{equation}
where $p_{\!_{R\!\gamma}}$ is the probability of reaction in the
reactive state [see eqs. (\ref{cage2}) and (\ref{dif2})] and
$p_s^{}$ is statistical factor, representing the relative number of
reactive spin states: for $D\!-\!D$ and $T\!-\!D$ pairs we get
$p_s^{} = 1/4$ and $p_s^{} = 1/3$, respectively. In eq.
(\ref{dis1a}) we neglected small contribution of quantum transitions
to the reaction rate.

Formulas (\ref{dis1}) and (\ref{dis1a}) are fairly suitable for
studying the efficiency of the ESP generation in different types of
bulk processes. The explicit expressions for the ESP are, naturally,
determined by the particular kinetic scheme of the process, and
should be discussed as applied to the interpretation of specific
experiments.

\subsubsection{Effect of attractive interparticle interaction}

The proposed theory is based on the assumption that particles
$X_a^{}$ and $X_a^{}$ [see eq. (\ref{gen0})] undergo free stochastic
motion. In reality, however, the ESP can be strongly affected by the
interparticle interaction $U(r) = k_{B}^{}T u(r)$.

Especially strong effect of the interaction is expected in
electron-hole ($e-h$) recombination, kinetics of which is known to
be markedly influenced by the Coulomb $e-h$ interaction. Significant
interaction effect can also be observed in some $T\!-\!D$ pairs, in
which $D$-particle is the electron ($e$) or hole ($h$).\cite{Prig}
For these pairs $U (r) \sim -\frac{1}{2}\alpha_{\!_T} E^2(r) $,
where $\alpha_{\!_T}$ is a polarizability of $T$ state and $E(r) =
e/(\varepsilon r^2_{})$ is the strength of the Coulomb field of $e$
(or $h$) particle in the semiconductor with dielectric constant
$\varepsilon$.

In the presence of the well-type attractive potential $U(r)$ the ESP
${\cal P}_{c}^{}$  can approximately be estimated within the simple
kinetic scheme, taking into account the effect of the cage within
the well,\cite{Shu5,Shu6} which hereafter is denoted as $U$-cage.
The size $R_c^{}$ of the $U$-cage, determined by the relation
$|U(R_c^{})| = k_B^{}T$, is expected to be significantly larger than
the size $r_c^{}$ of the primary cage, considered above in the cage
model: $R_c^{} \gg r_c^{} \sim \lambda$.

The kinetic scheme of the process in the $U$-cage (in the well) can
also be represented as (\ref{gen0}), but with the reaction (${\cal
W}_r^{}$) and dissociation (${\cal W}_d^{}$) rates\cite{Shu5,Shu6}
\begin{equation} \label{dis4}
{\cal W}_r^{} = {\cal K}_r^{}/{\cal V}_{\!_U} \;\;\mbox{and}
\;\; {\cal W}_d^{} = (K_0^{}/{\cal V}_{\!_U})e^{-u_a{}},
\end{equation}
in which $u_a^{} = U_a^{}/(k_{\!_B}T)$ is activation energy of
dissociation (the well depth), $K_0^{} = 4\pi R_c^{} D_{\!p}^{}$ is
the bimolecular rate of capture into the well, and ${\cal V}_{\!_U}$
is the volume of the $U$-cage. The volume can be estimated as ${\cal
V}_{\!_U} \approx (4 \pi/3)R_{\!_U}^{3}$, where $R_{\!_U}^{}$ is the
radius of the thermal distribution within the well, i.e.
$U(R_{\!_U}^{}) \approx U_b^{} + k_{\!_B}^{} T$, with $U_b^{} =
-U_a^{}$ being the energy of the bottom of the well.

In the $U$-cage the ESP generation is described by the rate
\begin{equation} \label{dis4a}
{\cal W}_c^{} = {\cal P}{\cal W}_0^{},
\; \;\mbox{where} \;\;{\cal W}_0^{} = k_0^{}/{\cal V}_{\!_U}.
\end{equation}

The kinetic scheme (\ref{gen0}) predicts exponential kinetics of the
decay of $U$-cage population, but with different decay rates for
different spin states of pairs. In particular, for non-reactive
states ($T$ and $Q$ states for $D\!-\!D$ and $T\!-\!D$ pairs,
respectively), whose depopulation kinetics determine the ESP
generation, the decay rate is ${\cal W}_d$. In such a case the
$U$-cage affected ESP ${\cal P}_{c}^{}$ is written as
\begin{equation} \label{dis5}
{\cal P}_{c}^{} = {\cal P}({\cal W}_0^{}/{\cal W}_d^{})
 = {\cal P}(r_c^{}/R_c^{})e^{u_a}
\end{equation}
(recall that for $T\!-\!D$ system ${\cal P}\equiv {\cal
P}_{\!*}^{}$). Similarly the $U$-cage affected ESP generation rate
${\cal K}_{c}^{}$ is given by ${\cal K}_{c}^{} = {\cal
P}_{c}^{}K_0^{} = {\cal K}e^{u_a}$. As for the decay rate of
reactive states, it is equal to ${\cal W}_r + {\cal W}_d$ and,
therefore, the $U$-cage affected reaction rate is represented as
${\cal K}_{r_c}^{} = [{\cal W}_r^{} /({\cal W}_r + {\cal W}_d)]K_c$.

It is seen from eq. (\ref{dis5}) that the attractive interparticle
interaction can result in the strong increase of the ESP.

Noteworthy is that formulas (\ref{dis5}) are obtained for not very
deep wells, when $\eta = {\cal W}_c^{}/{\cal W}_d^{} < 1$ and one
can neglect the contribution of quantum transitions to the reaction
rate ${\cal W}_r^{}$. To find ${\cal P}_{c}^{}$ and ${\cal
K}_{c}^{}$ in the opposite limit $\eta > 1$, one has to solve
complicated kinetic equations, in general. Some simple estimations
can, nevertheless, be made in the realistic case of strong $J(r)$,
large Zeeman splitting $\omega_{\gamma}^{} > 1$, and high reactivity
(when $p_{\!_{R_\gamma}} \approx 1)$. In this case only quantum
transitions between non-reactive and reactive states, localized in
crossing regions, contribute the ESP. These contributions can be
easily evaluated for both $D\!-\!D$ and $T\!-\!D$ reactions:

a) For $D\!-\!D$ system only $T_{\!_-} \to S$ transitions contribute
to the ESP so that ${\cal P}_{c}^{}$ can be written as
\begin{equation} \label{dis6}
{\cal P}_{c}^{} = \mbox{$\frac{1}{4}$}
{\cal W}_{\!_{ST_{\!_-}}}/({\cal W}_d^{} +
{\cal W}_{\!_{ST_{\!_-}}}\!),
\end{equation}
where ${\cal W}_{\!_{ST_{-}}} = {\cal W}_{\!_{S \leftarrow T_{-}}} =
4{\cal P}{\cal W}_0^{}$ is the transition rate, whose value is
obtained by comparing the expression (\ref{dis6}) with eq.
(\ref{dis5}) in the limit ${\cal W}_{\!_{ST_{-}}} \ll {\cal
W}_d^{}$.

b) For $T\!-\!D$ system similar analysis leads to formula
\begin{eqnarray} \label{dis6a}
{\cal P}_{c}^{} \equiv {\cal P}_{*_c}^{} &=& \mbox{$\frac{1}{6}$}
[{\cal W}_{2}^{}/({\cal W}_d^{} + {\cal W}_{2}^{}) +
{\cal W}_{\!_{-}}/
({\cal W}_d^{} + {\cal W}_{\!_+})]\quad\;\nonumber\\
&=&({\cal P}_{\!*}^{}{\cal W}_0^{})\big/
\big[{\cal W}_d^{} +
(9/2){\cal P}_{\!*}^{} {\cal W}_0^{}\big]
\end{eqnarray}
in which ${\cal W}_{\!_{\pm}} =  {\cal W}_{3}^{} \pm {\cal W}_{1}^{}
$, with ${\cal W}_{1}^{} \equiv {\cal W}_{_{D_{\!-1/2} \leftarrow
Q_{-3/2}}} = \frac{3}{2}{\cal P}_{\!*}^{} {\cal W}_0^{}$ and ${\cal
W}_{3}^{} \equiv {\cal W}_{_{D_{\!+1/2} \leftarrow Q_{-3/2}}} = 2
{\cal W}_{1}^{}$, and ${\cal W}_{2}^{} \equiv {\cal W}_{_{D_{\!+1/2}
\leftarrow Q_{-1/2}}} = 3 {\cal W}_{1}^{}$. Similarly to the case of
$D\!-\!D$ system the expression of the rates ${\cal W}_{j}^{}, \: (j
= 1 - 3),$ in terms of the ESP ${\cal P} \equiv {\cal P}_{\!*}^{}$
is found by comparison of eq. (\ref{dis6a}) with eq. (\ref{dis5})
(for ${\cal P}_{\!c}^{} \equiv {\cal P}_{\!*_c}^{}$) at ${\cal
W}_{1}^{} \ll {\cal W}_d^{}$.

Formulas (\ref{dis6}) and (\ref{dis6a}), looking quite natural in
the diffusion model, need some comments as applied to the cage
model. These expression imply that within the primary cage (at $r
\lesssim r_c^{}$) a large variety of relative positions $r$ of
particles, including those in the term-crossing region, are
accessible. Note that possible migration of particles over these
positions can result in $V (t)$-fluctuations and, thus, to some
change of $w_c^{}$ (see Sec. IIIB.1 and IVB.1). However, in the
considered case of strong $J(r)$ and large $\omega_{\gamma}^{}$ this
change does not lead to any change of the ESP.\cite{Shu1,Shu2,Shu3}
Noteworthy is also that the above-mentioned large variety of $r$ in
primary cages can be realized as a result of reencounters of
reacting particles at different points in the $U$-cage. In such a
case the ESP is quite properly described by the original eqs.
(\ref{dis6}), (\ref{dis6a}).

\subsubsection{Magnitude of the ESP}

As we have already noted above, the dependence of the $D\!-\!D$ ESP
${\cal P} (\omega_{\gamma}^{}) \sim \Phi_{\mu}^{\chi}
(\omega_{\gamma}^{})$ on the dimensionless Zeeman splitting
$\omega_{\gamma}^{} = \Omega_0^{}/w_{\gamma}^{}, (\gamma = c, f),$
is non-monotonic for both HFI and AZI mechanisms of ESP generation
(see Sec. IIIA) and both models of relative motion: the cage
($\gamma = c$) and free diffusion ($\gamma = f$) [except the case
$\mu = Z, \,\chi = s$ (\ref{pro2})]. Similar non-monotonic behavior
of ${\cal P}_{\!*}^{} (\omega_{\gamma}^{})$ is observed in
$T\!-\!D$-quenching [eq. (\ref{td9})]. In all these cases the
maximum value ${\cal P}_{\!m}^{} = {\cal P} (\omega_{\gamma_m}^{})$
[or ${\cal P}_{\!m}^{} = {\cal P}_{*} (\omega_{\gamma_m}^{})$ for
$T\!-\!D$ process] is located at $\omega_{\gamma_m}^{} \sim 1$.

Such a behavior of ${\cal P} (\omega_{\gamma}^{})$ allows one to
simplify the analysis of the ESP by restricting the discussion to
maximal values ${\cal P}_{\!m}^{}$ of the effect. In what follows,
for certainty, we will estimate ${\cal P}$ at $\omega_{\gamma}^{} =
1$, for which $\Phi_{\mu}^{\chi} (1) \sim [\Phi_{\!_H}^{\chi} (1) +
2\Phi_{\!_H}^{\chi} (2)] \sim 1$, so that the maximal values ${\cal
P}_{\!m}^{}$ for $D\!-\!D$ and $T\!-\!D$ processes can be written as
\begin{equation} \label{dis7}
{\cal P}_{\!m}^{} \sim p_{\!_{R\!_\gamma}}^{}\! \langle
v_{\!\mu}^2\rangle \bar P_{\!\chi}, \;\; (\mu = H, Z, T
\;\:\mbox{and}\;\: \chi = s, w),
\end{equation}
In this formula $p_{\!_{R\!_\gamma}}^{}$ is the probability of
reaction in reactive spin states [see eqs. (\ref{cage2}) and
(\ref{dif2})]. The parameter $\langle v_{\!\mu}^2\rangle$ is the
dimensionless coupling of terms of $H_0^{}(r)$ [eqs. (\ref{pro1a})
and (\ref{td9}). As for $\bar P_{\!\chi}$, it describes the
dependence of the ESP on the strength of $J (r)$ ($\chi = s$ and
$\chi = w$ for strong and weak $J$, respectively). Formulas for
$\bar P_{\!\chi}$, different in two considered models of motion, are
presented in eqs. (\ref{pro3a}), (\ref{pro3b}) and (\ref{pro6b}).

Of certain interest is the above-mentioned case of AZI mechanism
($\mu = Z$) and strong $J(r)$ ($\chi = s$), in which
$\Phi_{\!_Z}^{s} (\omega)$ is a monotonically increasing function.
Below it will be analyzed more thoroughly. Here we only note that in
this case at large $\omega \sim B$ the ESP can be very large.

Now we will estimate the parameters in eq. (\ref{dis7}):

1. The largest ESP is, naturally, expected in the case of strong
reactivity resulting in a high reaction probability
$p_{\!_{R\!_\gamma}}^{}$. In what follows we will assume that
$p_{\!_{R\!_\gamma}}^{} \sim 1$.

2. The parameter $\langle v_{\!\mu}^2\rangle$ ($\mu = H, Z, T$) is
suggested to be small: $\langle v_{\!\mu}^2\rangle  < 1$, to ensure
the validity of the applied perturbation approximation (in $V$).
Noteworthy is that for AZI mechanism ($\mu = Z$) the validity of
this approximation implies also not very large value of the magnetic
field $B$ (to be sure that the value of $\langle V_{\!_Z}^2\rangle
/w_c^2$ is small).

3. The parameter $\bar P_{\!\chi}$ [as it follows from eqs.
(\ref{pro3a}), (\ref{pro3b}) and (\ref{pro6b})] is smaller than $1$,
but is not very small. The fact is that, in reality, $w_e^{} \sim
w_d^{} \sim w_{\!f}^{}$ and $D_p^{} \sim \lambda^2/w_d^{} \sim
d^2/w_d^{}$, so that the small value of $\bar P_{\!\chi}$ results
from the relation $ \bar P_{\!\chi} \sim (\alpha d)^{-1}$, in which
$\alpha d \lesssim 10$.


Combining the estimations of these three parameters we can obtain
the ESP, but to calculate it more accurately we need to specify the
process under study.

For example, let us consider the cage model prediction for the ESP
${\cal P}\equiv{\cal P}_{*}^{}$ of $D^{*}$ particles in $T\!-\!D$
quenching. The value of the parameter $\langle v_{\!_T}^2\rangle$
for the $T\!-\!D$ system is determined by the ZFS constant
$D_{\!_T}^{}$ and $w_c^{} \sim w_h^{}$. For realistic values
$D_{\!_T}^{} = 3 \cdot 10^{9}s^{-1}$ and $w_h^{} = 10^9 s^{-1}$ one
obtains  $\langle v_{\!_T}^2\rangle =
\frac{2}{45}(D_{\!_T}^{2}/w_c^2) \approx 0.4$. Taking into account
the above estimations $p_{\!_{R\!_\gamma}}^{} \sim 1$ and $(\alpha
d)^{-1} \gtrsim 0.1$ we get (for $\omega_c^{} = 1$) the value ${\cal
P}_{\!*_m}^{} \gtrsim 4 \cdot 10^{-2}$, much larger than the thermal
ESP ${\cal P}_{\!*_{th}}^{} \approx 1.4 \cdot 10^{-3}$ (at room
temperature).

This value of ${\cal P}_{\!*_m}^{}$ is obtained in the case of
strong exchange interaction ($\chi = s$), predicting largest ESP. Of
course, in the opposite limit ($\chi = w$) the ESP is smaller with
the factor $J_0^{}/w_{\gamma}^{} \ll 1$ [see eq. (\ref{pro6b})].

The attractive $T\!-\!D$ interaction can result in the strong
increase of the $U$-cage affected ESP ${\cal P}_{\!*_c}^{}$: ${\cal
P}_{*_c}^{} \sim {\cal P}_{\!*}^{} e^{u_a^{}}$, where $u_a^{}$ is
the depth of the potential well ($U$-cage), as it follows from eq.
(\ref{dis5}). In particular, even in the case of fairly shallow
attractive well of the interaction potential, corresponding to
$u_a^{} =2.0\, (k_B T)$, this interaction gives rise to the
significant (about one order of magnitude) increase of the ESP
${\cal P}_{\!*_e}^{}$. Taking into account the above-obtained
estimation of ${\cal P}_{\!*}^{}$, we expect, that the interaction
affected $T\!-\!D$ ESP can be fairly large: ${\cal P}_{\!*_e}^{}
\sim 0.1$.

Similar estimations can also be made for the $D\!-\!D$ ESP. Because
of this similarity we are not going to do them here, but restrict
ourselves to some comments on the AZI mechanism of the ESP ($\mu =
Z$). The fact is that values of $\langle v_{\!_Z}^2 \rangle$ [see
eq. (\ref{pro1a})] typical for organic semiconductors are very
small: $\langle v_{\!_Z}^2 \rangle \lesssim 10^{-5}$ [due to small
$g'\!\!:\!g'$ (\ref{ehr8})], and for not very strong magnetic fields
$B$, corresponding to $\omega_{\!\gamma}^{} \lesssim 10^{3}$, the
ESP is fairly small: ${\cal P} \lesssim 10^{-3}$. It can be
significantly larger, however, in the presence of the attractive
interaction ($U$-cage), as it follows from above relations. Note
also, that in some organic semiconductors, doped with heavy atoms,
the Zeeman interaction can be strongly anisotropic,\cite{Sw}
resulting in fairly large $\langle v_{\!_Z}^2 \rangle \gtrsim
10^{-4}$ and, thus, in the ESP much larger than the thermal one.

Concluding our analysis of the $U$-cage effect on the ESP, we recall
that the applied formulas (\ref{dis4}) and (\ref{dis5}) are valid
for relatively shallow wells, when $\eta = ({\cal P}{\cal
W}_0^{})/{\cal W}_d < 1$. For deeper wells, for which $\eta > 1$, a
fairly large $U$-cage effect is expected, though the calculation of
the ESP in this limit is a rather complicated problem, in general.
Fortunately, in the realistic case of strong $J(r)$ and large Zeeman
splitting $\omega_{\gamma}^{}$ it is simplified, as has been shown
above [eqs. (\ref{dis6}) and (\ref{dis6a})]. Obtained formulas
predict that in the limit $\eta \gg 1$ the ESP is very large: ${\cal
P}_{\!c}^{} (\eta \gg 1) = 1/4$ and ${\cal P}_{\!*_c}^{} (\eta \gg
1) = 2/9$ for $D\!-\!D$ and $T\!-\!D$ processes, respectively.

Naturally, all above discussed effects manifest themselves in the
rates of ESP generation in corresponding bulk processes as well. We
are not going to analyze this manifestation. Such an analysis would
require the specification of the process under study, which should,
perhaps, be done as applied to particular experiments.

Here we will only restrict ourselves to a general remark on the
relation between geminate and bulk ESP. Note, that the ESP can be
analyzed in terms of the ratio of the ESP generation (${\cal K}$)
and reaction (${\cal K}_r^{}$) rates: $R_{\!_{\cal K}} = {\cal
K}/{\cal K}_r^{} = (p_{\!_S}p_{\!_{R_{\!\gamma}}})^{-1} {\cal P}$.
Similar relation is valid in the presence of the attractive
potential [see eqs. (\ref{dis5})]. All these relations show that the
analysis of the ESP generation in bulk processes provides the
information identical to that obtained by studying the geminate ESP.

\section{Conclusions}

In this work we have analyzed the specific features of the ESP of
charge carriers (electrons and holes) in $D\!-\!D$ recombination and
$T\!-\!D$ quenching in disordered semiconductors. Simple expressions
for the ESP, generated in both processes, are derived for two models
of stochastic relative motion of particles: cage and free diffusion.
Both models predict fairly large ESP in geminate $D\!-\!D$ and
$T\!-\!D$ processes (much larger than the thermal ESP at room
temperatures). Especially large ESP is expected in the presence of
attraction between particles. Naturally, high efficiency of the ESP
generation is predicted in corresponding bulk processes as well.

The mechanism , considered in our work, is based on the assumption
that the ESP results from quantum transitions between the terms of
the spin Hamiltonian of reacting pairs at short distances ($r \sim
d$) near the term crossing regions (as applied to $D\!-\!D$
recombination it is called $ST_{\!_{-}}$-mechanism\cite{St}).

It is worth noting, however, that there is another pair mechanism,
well known in the theory of the $D\!-\!D$ ESP as $ST_{\!{0}}$
mechanism.\cite{St}. It also contributes to the ESP of $D$ particles
in the case $\Delta g = \bar g_a^{} - \bar g_b^{} \neq 0$ [see eq.
(\ref{ehr9})], but according to the
estimations\cite{Goud,Shu1,Shu2,Shu3}, for typical parameters of
systems considered this contribution is much smaller than those
considered in our work.

\section{Acknowledgements}\, The work was partially supported by the
Russian Foundation for Basic Research.


\end{document}